\documentclass[prl,twocolumn,showpacs]{revtex4}

\usepackage{graphicx}
\usepackage{dcolumn}
\usepackage{amsmath}
\usepackage{times}

\begin{document}
\title{Phonon-Metamorphosis in Ferromagnetic Manganite Films:\\
Probing the Evolution of an Inhomogeneous State}

\author{Ch.~Hartinger}
\affiliation{EP V, Center for Electronic Correlations and
Magnetism, University of Augsburg, 86135 Augsburg, Germany}

\author{F.~Mayr}
\affiliation{EP V, Center for Electronic Correlations and
Magnetism, University of Augsburg, 86135 Augsburg, Germany}

\author{A.~Loidl}
\affiliation{EP V, Center for Electronic Correlations and
Magnetism, University of Augsburg, 86135 Augsburg, Germany}

\author{T.~Kopp}
\affiliation{EP VI, Center for Electronic Correlations and
Magnetism, University of Augsburg, 86135 Augsburg, Germany}

\date{\today}

\begin{abstract}
The analysis of phonon anomalies provides valuable information
about the cooperative dynamics of lattice, spin and charge degrees
of freedom. Significant is the anomalous temperature dependence of
the external modes observed in La$_{2/3}$Sr$_{1/3}$MnO$_{3}$
(LSMO) films. The two external modes merge close to the
ferromagnetic to paramagnetic transition at $T_C$ and, moreover,
two new modes evolve in this temperature range with strong
resonances at slightly higher frequencies. We propose that this
observed phonon metamorphosis probes the inhomogeneous Jahn-Teller
distortion, manifest on the temperature scale $T_C$. The analysis
is based on the first observation of all eight phonon modes in the
metallic phase of LSMO and on susceptibility measurements which
identify a Griffiths-like phase.

\end{abstract}


\maketitle The colossal magnetoresistance (CMR) in the manganites
is a paradigm
of the cooperative dynamics of charge, spin and lattice degrees of
freedom (DOF). Whereas the interplay of charge and spin DOF was
realized early on with the discovery of the double exchange
mechanism (DE), the role of the lattice DOF was appreciated only
recently:  Millis {\it et al.\/}\cite{Millis95} argued that the DE
alone cannot explain the resistivity data of
La$_{1-x}$Sr$_{x}$MnO$_{3}$. They concluded that a sufficiently
strong electron-phonon coupling is responsible for polaronic
effects which control the electronic kinetic energy and the
resistance peak near the ferromagnetic to paramagnetic (FM-PM)
transition.

A new perspective was introduced through the notion that the
formation of an inhomogeneous state in the vicinity of the
transition plays a key role in the CMR phenomenon
\cite{Uehara99,Moreo99,Dagotto01}. It not only signifies the
competition between the different thermodynamic states which may
result from the cooperative dynamics of the considered DOF, but it
also emphasizes the relevance of disorder which, through doping,
is present even in all CMR manganites. In this context, the idea
that ``CMR is a Griffiths singularity''
\cite{Salamon02,Griffiths69} has been advanced.

Experimental evidence for the presence of inhomogeneous states in
the temperature range where the CMR is observed has been reported
on many occasions and has been reviewed before (see for example
\cite{DagottoBook}). It should be emphasized that the several
investigated inhomogeneous states are of different nature although
they probably have a common origin which is not yet fully
understood. Prominent examples of such investigations are:
scanning tunnelling spectroscopy \cite{Faeth99,Samwer03} which
probes the surface states, susceptibility measurements
\cite{Salamon02} which display non-Curie-Weiss behavior and refer
to competing magnetic states, and neutron-diffraction studies
\cite{Louca99} which examine the local Jahn-Teller (JT)
distortions related to a microscopic charge distribution.

Most of the previous research focussed on
La$_{1-x}$Ca$_{x}$MnO$_{3}$, as the CMR effect is strongest in the
Ca-doped compounds, at $x\simeq 1/3$. However, since for these
compounds the ferromagnetic to paramagnetic transition ($T_C$)
nearly coincides with the metal-insulator transition ($T_{\rm
MIT}$), it is of particular relevance to investigate
La$_{1-x}$Sr$_{x}$MnO$_{3}$ where the two temperature scales
are different. In this work, we address the observability of the
inhomogeneous bulk state through far infrared (FIR) optical
spectroscopy in La$_{2/3}$Sr$_{1/3}$MnO$_{3}$ (LSMO) films. We
associate phonon anomalies with the formation of the inhomogeneous
state. To our knowledge, anomalies of the phononic excitations in
LSMO have not been analyzed in the vicinity of the magnetic
transition. Probing  the inhomogeneous state with phonons should
provide insight into the temperature-dependent evolution of
electronic correlations which couple to the local lattice
structure.

The origin of phononic anomalies may be traced (i)~to a strong
coupling of  phonon modes with an electronic continuum but also
(ii)~to a modification of the local symmetry, e.g., due to an
alteration of the JT distortion. Mechanism~(i), which leads to an
asymmetric ``Fano lineshape'' of distinct phonons, is observed in
La$_{2/3}$Ca$_{1/3}$MnO$_{3}$ films \cite{Hartinger04b}. In LSMO
films the electron-phonon coupling is much weaker, as evidenced by
the polaronic excitations \cite{Hartinger04a}, and correspondingly
the asymmetry of the phonon absorption lines is too small to be
observable. However, the latter effect (ii), which we denote as
``metamorphosis of phonons'', is seen in our LSMO films. The
metamorphosis not only includes a temperature dependent frequency
shift but, more notably, the merging of phonon lines and the
evolution of new resonances. We observe a transfer of spectral
weight from the pronounced phonon modes at low temperature into
the ``new modes'' at high temperature. This metamorphosis is
controlled by the magnetic temperature scale $T_C$.

The study of thin films is indispensable in order to resolve the full
set of phonon modes even in the metallic low-temperature phase.
Enhanced screening in the single crystals
suppresses the phonon resonances wherefore there are no reports on
the infrared active phonons in
La$_{2/3}$Sr$_{1/3}$MnO$_{3}$
available in the literature. Apart from rendering the phonons
observable,  thin films also serve the purpose of extending the
inhomogeneous state to a wider temperature range as lattice strain
and imperfections support the Griffiths-like phase.

Measurements were performed on a single crystal and two thin films
of La$_{2/3}$Sr$_{1/3}$MnO$_{3}$, which were grown onto
(LaAlO$_{3}$)$_{0.3}$(Sr$_{2}$AlTaO$_{5}$)$_{0.7}$. The film
thickness was  300~nm for LSMO~\#1 and 400~nm for LSMO~\#2. X-ray
analysis revealed a rhombohedral structure for all samples. The
reflectivity measurements were carried out between 50~cm$^{-1}$
and 40000~cm$^{-1}$, using the Fourier transform spectrometers
Bruker IFS 113v and IFS 66v/S. Careful determination of the
optical conductivity $\sigma$ was achieved by using results of a
submillimeter interferometer for the low-energy range in the
Kramers-Kronig analysis. Detailed information are published
elsewhere \cite{Hartinger04b}. The DC resistivity was measured
using a four-point configuration. Magnetization measurements were
performed with a  commercial Quantum Design SQUID magnetometer.

\begin{figure}[t]
\vspace{0mm} \centering
\includegraphics[width=.47\textwidth,clip,angle=0]{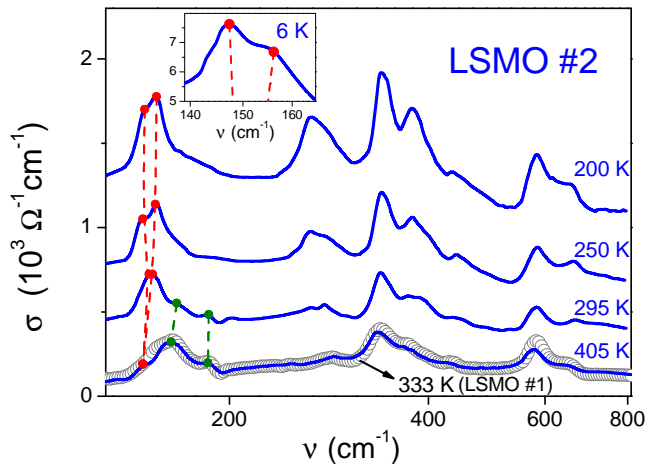}
\caption[]{\label{Sr-sigma-T} Temperature-dependent optical
conductivity ($\sigma$) spectra of LSMO~\#2 (solid lines), with a
spectral resolution $\sim 1$~cm$^{-1}$ on a logarithmic frequency
scale. We found similar spectra for LSMO~\#1 (open circles),
except for a reduction of their absolute values. The vertical
lines connect the maxima of the external modes. Inset: At 6~K the
splitting of the external modes is strongest.}
\end{figure}

At low temperature long-range ferromagnetic ordering induces
the metallic state. With increasing temperature the FIR optical
conductivity $\sigma$ diminishes up to the metal-insulator
transition at $T_{\textrm{MIT}}$. Figure~\ref{Sr-sigma-T} displays
$\sigma$ in the energy range of the phononic excitation for
LSMO~\#2 (lines) and LSMO~\#1 (open circles). We identify eight
peaks of active transverse-optical (TO) phonon modes,
corresponding to all IR active vibrations from
R$\bar{3}$c-symmetry. In Tab.~\ref{tabmoden} we present a
comparison between the experimental values of the infrared-active
phonon frequencies of LSMO \#2 at $T=405$~K and the calculated
frequencies for pure rhombohedral  LaMnO$_{3}$ by Abrashev
\textit{et al.} \cite{Abrashev99}. The calculated positions for
the undoped compound should be taken as a rough estimate for LSMO
since the lattice constants and atomic positions are modified  by
doping. Several eigenfrequencies are known from neutron scattering
measurements, which are in good agreement with our
data~\cite{Reichardt99}.

\begin{table}[b]
\caption[]{Correspondence between calculated \cite{Abrashev99} and
measured phonon modes (LSMO~\#2) for the space group R$\bar{3}$c
(in cm$^{-1}$) at  $T=405$~K, and the values in parenthesis to
neutron scattering~\cite{Reichardt99} with
La$_{0.7}$Sr$_{0.3}$MnO$_{3}$ samples. \label{tabmoden}}
\begin{center}
\begin{tabular}{clccccllclcc} \hline \hline
 &\multicolumn{3}{c}{calculated}      &&    &  &\multicolumn{3}{c}{measured}     &\\
 &$A_{2u}$ &  & $E_{u}$  &&    assignment &  &$A_{2u}$ &  & $E_{u}$  &\\ \hline
 &         &  & 317      && vibration (Mn)& &         &  & 373      & \\
 &162      &  & 180      && external& &150      &  & 150      & \\
 &310      &  & 357      && bending&  &338  (336)    &  & 436  (424)    &\\
 &641      &  & 642      && stretching&  &580 (576)     &  & 650      &\\
 &         &  & 240      && torsional&  &         &  & 285      &\\  \hline \hline
\end{tabular}
\end{center}
\end{table}

A particular temperature dependence of the mode frequency is
observed for the external modes. With increasing temperature the
two external modes approach each other and the merged modes shift
to lower frequency above $T_{C}\approx345$~K
(cf.~Fig.~\ref{Sr-sigma-T}). A degeneracy of modes corresponds to
a higher lattice symmetry. Moreover, at approximately room
temperature, two new modes appear, pointing to a lower lattice
symmetry. The new modes gain weight for temperatures close to
$T_C$ so that they dominate the group of external modes above
$T_C$. This phenomenon of phonon metamorphosis is observable in
both films, in  LSMO~\#2 as well as LSMO~\#1 (cf.\ open circles in
Fig.~\ref{Sr-sigma-T}). The groups of stretching and bending modes
are weakly affected by the FM-PM transition except for the torsional
mode which is shifted to higher frequencies.

\begin{figure}[t]
\centering
\includegraphics[width=.5\textwidth,clip,angle=0]{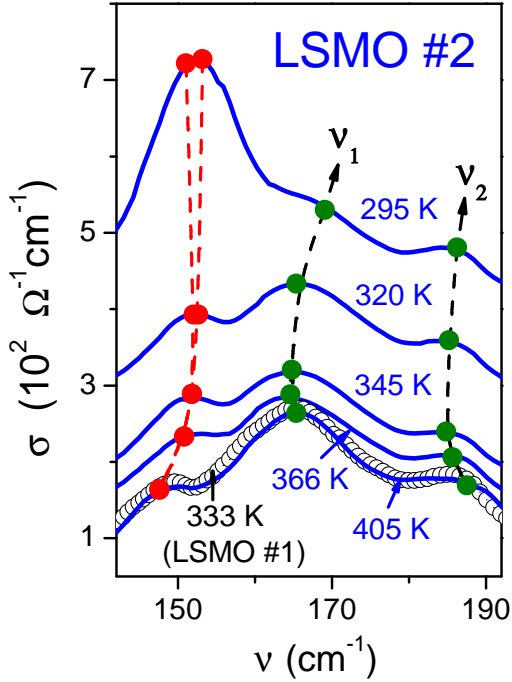}
\vspace{0mm} \caption[]{\label{HT_Mode_Sr} Metamorphosis of phonon
modes: optical conductivity
$\sigma$ of the external modes and the new modes ($\nu_{1}$
$\nu_{2}$). The latter modes are absent at low temperature.
With increasing temperature, spectral weight
is transferred from the low-energy external group to the new modes.}
\end{figure}

\begin{figure}[b]
\centering
\includegraphics[width=.47\textwidth,clip,angle=0]{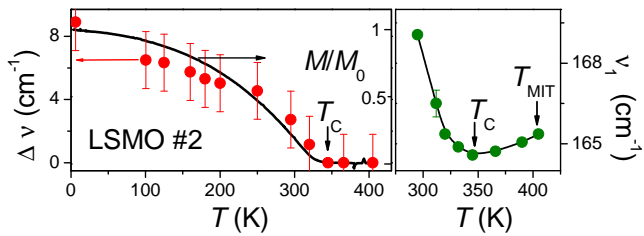}
\vspace{0mm} \caption[]{\label{HT_nu_Sr}  Left panel: Splitting
$\Delta\nu$ between the eigenfrequencies of the two external modes
at various temperatures (full circles). The solid line is the
temperature dependent magnetization of  LSMO~\#2. Right panel:
Temperature dependence of the frequency of the new mode $\nu_{1}$.
At $T_{C}$ the direction of the shift is reversed.}
\end{figure}

The external group on an enlarged scale is displayed in
Fig.~\ref{HT_Mode_Sr}. The two new modes at
approximately 169 and 186~cm$^{-1}$ are referred to as $\nu_{1}$
and $\nu_{2}$, respectively. The selected temperatures fully cover
the interval in which both, the magnetic and metal-to-insulator
transitions, take place ($T_C\approx345$~K and $T_{\rm
MIT}\approx401$~K). The presumption that the evolution of these
modes is controlled by the spin DOF is substantiated by the
temperature analysis of the mode frequencies. The temperature
dependence of the splitting of the low frequency pair of modes
$\Delta\nu$ is presented in Fig.~\ref{HT_nu_Sr} (left panel, full
circles). The solid line is the normalized magnetization
($M(T)/M(0)$), displayed in this left panel. Although no
theoretical modelling is available to relate magnetization and
mode splitting quantitatively, the observation that $M$ and
$\Delta\nu$ drop on the same temperature scale $T_C$ to zero
suggests a correlation between spin polarization and these phonon
modes. The same observation is true  for the temperature
dependence of the mode frequencies $\nu_{1}$ and $\nu_{2}$. The
temperature scale is clearly $T_C$, and not $T_{\rm MIT}$, as
shown for $\nu_{1}(T)$ in the right panel of Fig.~\ref{HT_nu_Sr}.
Neutron scattering measurements \cite{Martin96,Mellergard00} show
that the structural distortion does not change at $T_C$
and thereby confirm that neither the
lattice structure nor the thermal expansion of the lattice can be
attributed to the temperature behavior of the external modes in
the vicinity of $T_{C}$. It is suggestive to relate the observed
phonon metamorphosis to the formation of an inhomogeneous state,
as this state may support two types of external modes.

\begin{figure}[htb]
\centering
\includegraphics[width=.5\textwidth,clip,angle=0]{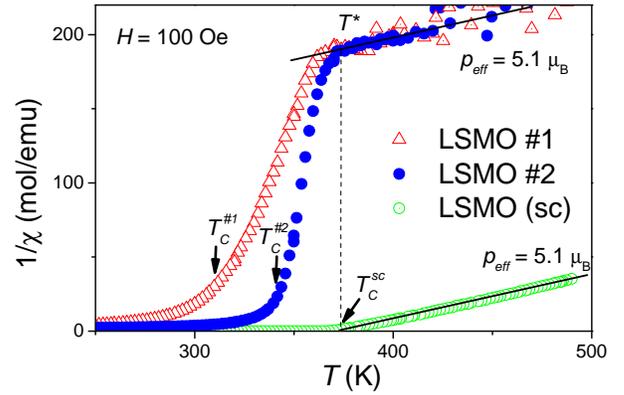}
\vspace{0mm} \caption[]{\label{Griffiths-Phase} Temperature
dependence of the low-field inverse magnetic susceptibility for
three samples (LSMO (sc): single crystal,  LSMO \#2 and LSMO \#1).
Arrows indicate $T_{C}$ from magnetization measurements. The
effective magnetic moment  $p_{\rm eff}$ in the paramagnetic phase
above $T^*$ is $5.1 \mu_B$ for all samples \cite{comment2}.}
\end{figure}

An inhomogeneous magnetic state may be identified from
susceptibility measurements
in low magnetic fields. Salamon and Chun \cite{Salamon03} could
not detect the onset of a Griffiths-like phase for
La$_{0.7}$Sr$_{0.3}$MnO$_{3}$ single crystals. However, thin films
are intrinsically more inhomogeneous and the question arises if
they display Griffiths singularities. In fact, our measurements of
the susceptibility $\chi$ in LSMO~\#2 ($400$~nm film) and LSMO~\#1
($300$~nm film) in Fig.~\ref{Griffiths-Phase} support
such a scenario. Whereas the single crystal with the same
Sr-concentration shows the conventional Curie-Weiss behavior with
an effective magnetic moment $p_{\rm eff}=5.1 \mu_B$ (open
circles), the films display a sharp downturn in $\chi^{-1}(T)$
at a temperature scale $T^* \sim 370$~K above their respective
Curie temperatures $T_C^{\#1} = 310$~K and $T_C^{\#2} = 345$~K.
The onset $T^*$ is identical for both films and
corresponds to the Curie temperature $T_C\sim 370$~K of the single crystal.
Such a behavior may be interpreted in terms of a Griffiths-like
phase, where the enhanced disorder in the films, in comparison to
the single crystal, accounts for the reduced $T_C$ and the
formation of inhomogeneous magnetic states at $T^* \sim T_C^{sc}$
\cite{Griffiths69}.

Whereas the non-Curie-Weiss behavior of the susceptibility points
to the existence of spin clusters in the films, it does not
predict the temperature at which an inhomogeneous state first
develops. Probably ``electronic phase separation between phases
with different densities on a nano scale'' \cite{Dagotto01} is
established  before the disorder-induced Griffiths-like phase for
the spin system sets in. Even if the underlying model were known,
it would be difficult to make conclusive predictions as the phase
separation in these systems seems to be dynamic \cite{Heffner00}.
For a better characterization of the bulk inhomogeneous state in
the full temperature range up to the FM-PM transition we propose
to analyze the evolution of the considered phonon modes in more
detail, specifically their magnetic field dependence should be
addressed in future work. We  conclude the paper with a discussion
of the origin of the new modes.

We suggest that the observed phonon metamorphosis is to be
understood in the context of a {\it T}-dependent,
\textit{inhomogeneous dynamical Jahn-Teller} distortion (IDJT).
The IDJT is probed by distinct phonon excitations which ``sense''
the inhomogeneities as soon as the charge carriers slow down with
increasing temperature and the lattice DOF experience a crossover
from adiabatic to nonadiabatic behavior.

Louca and Egami \cite{Louca99}  investigated deviations of the
local structure from the average crystal structure  in
La$_{1-x}$Sr$_{x}$MnO$_{3}$ by pulsed neutron scattering with pair
distribution function analysis (PDF) and Shibata {\it et al.\/}
\cite{Shibata03} with x-ray absorption fine structure (XAFS).
Although their findings differ for the metallic phase
quantitatively, both groups identify a temperature-dependent,
inhomogeneous local lattice distortion. Louca and Egami introduced
the notation of an ``anti-JT distortion'' for the undistorted
sites (with Mn$^{4+}$) where the holes, i.e.\ the charge carriers,
reside at high temperature. With these findings the tentative
scenario for the merging of two phonons and the formation of new
modes is as follows:

At low temperatures, deep in the metallic phase, the dynamics of
the mobile charge carriers is fast with respect to the dynamics of
the lattice degrees of freedom: the lattice reacts on the average
charge carrier concentration, not on the instantaneous hole
positions. In this regime, only a weak JT-distortion is observed
and the external modes are split as seen for the 6~K line shape of
the inset of Fig.~\ref{Sr-sigma-T}. For increasing temperature,
the charge carrier dynamics is slower since the double exchange
mechanism is less effective due to fluctuations in the $t_{2g}$
spin directions, and the overall picture becomes more local. A
finite-size polaron emerges which is, according to Louca and
Egami, rather an anti-polaron where the lattice around the slower
holes (Mn$^{4+}$ sites) is locally less and less JT-distorted.
Correspondingly, the splitting of the external modes is reduced.
The weight of these phonon resonances decreases because sites
without anti-polarons (Mn$^{3+}$ sites) experience a strong
JT-distortion and
support external modes at a higher frequency (as
is well known from the undoped or weakly doped systems
\cite{Paolone00,Podobedov98,comment1}). These ``new external
modes'' at higher frequency also experience a larger splitting,
again consistent with the observations in undoped or weakly doped
systems. For even higher temperatures, close to $T_C$,  the
anti-polaron is reduced to a (JT undistorted) lattice site and the
dynamics of this hole is diffusive. The higher symmetry of these
sites supports only one external mode frequency whereas the
Mn$^{3+}$ sites with a full JT distortion now support two external
mode vibrations with higher spectral weight according to their
larger relative number. We refer to this situation as the
temperature dependent IDJT. The phase transition
scenario~\cite{Dagotto01} supports the IDJT. The basis for the
observation of the IDJT is a crossover from  adiabatic to
nonadiabatic behavior of the lattice dynamics.

To conclude, we want to emphasize the particular importance of the
external modes as an excellent phononic ``tool'' to investigate
this cooperative behavior of the lattice with spin and charge DOF.
These phonons have a weak dispersion \cite{Reichardt99} which
allows them to probe local distortions. Moreover, the
frequency shift of  the external modes is  strong with Sr-doping
because these modes correspond to a vibration of the La-Sr cage with respect to the
Mn-O octahedra. For this reason the low-energy modes of the external
group are well separated from the new modes at high temperature.
The disentanglement of the external modes in the electronic
inhomogeneous state is the prerequisite of the observed phonon
metamorphosis.

\begin{acknowledgments}
We acknowledge stimulating discussions with J.~Deisenhofer and
R.~Hackl.  The  research was supported in
part by  BMBF (13N6917A, 13N6918A) and by the  DFG via SFB 484 (Augsburg).
\end{acknowledgments}

%% \vskip -0.5cm


\begin{thebibliography}{1}
\bibitem{comment2} As a consequence of the imprecise mass determination of thin films
the scale of $\chi^{-1}(T)$ is subject to some uncertainty.
\bibitem{comment1} In La$_{0.95}$Sr$_{0.05}$MnO$_{3}$ the resonances of the external
    modes are at 169 and 180 cm$^{-1}$ for room temperature.
\end{thebibliography}
\end{document}